\documentclass[aps,prb,twocolumn,showpacs,floatfix,a4paper]{revtex4}
\usepackage{graphicx}
\usepackage{amsmath}

\begin{document}

\renewcommand{\theenumi}{\Roman{enumi}}
\renewcommand{\labelenumi}{\theenumi.}

\title{Interference effects in interacting quantum dots}
\author{Moshe Goldstein}
\author{Richard Berkovits}
\affiliation{The Minerva Center, Department of Physics, Bar-Ilan
University, Ramat-Gan 52900, Israel}

\begin{abstract}
In this paper we study the interplay between interference effects in
quantum dots (manifested through the appearance of Fano resonances
in the conductance), and interactions taken into account in the
self-consistent Hartree-Fock approximation.
In the non-interacting case we find that interference may lead to
the observation of more than one conductance peak per dot level as
a function of an applied gate voltage. This may explain recent
experimental findings, which were thought to be caused by interaction
effects.
For the interacting case we find a wide variety of different
interesting phenomena. These include both monotonous and
non-monotonous filling of the dot levels as a function of an applied
gate voltage, which may occur continuously or even
discontinuously. In many
cases a combination of the different effects can occur in the
same sample. The behavior of the population influences, in turn,
the conductance lineshape, causing broadening and asymmetry of
narrow peaks, and determining whether there will be a
zero transmission point.
We elucidate the essential role of the interference between the
dot levels in determining these outcomes. The effects of
finite temperatures on the results are also examined.
\end{abstract}


\maketitle

\section{\label{sec:intro} Introduction}

Transport in quantum dots has been the topic of an intense scrutiny
for more than twenty years (for a review, see, e.g.,
Ref.~\onlinecite{alhassid00}). However, most of both experimental
and theoretical studies were concentrated on either of the two
limits: (a) the limit of strong dot-lead coupling (``open dots''), in
which the discreteness of the dot's energy spectrum is completely
lost; (b) the limit of weak dot-lead coupling (``closed dots''), in
which, due to the Coulomb Blockade, each of the dot's levels creates
a well defined peak in the dependence of the dot's conductance on an
applied gate voltage.

In recent years, focus has shifted to the intermediate coupling case.
In this case, dot-lead coupling is weak enough so that the dot's
energy spectrum cannot be considered as a continuum, but there are
interference effects between different dot levels, manifested through
the appearance of Fano resonances in the dot's conductance
\cite{gores00, kobayashi02}.

Most the theoretical studies have so far concentrated on interference
effects in non-interacting dots \cite{clerk01, alhassid03, konig98}.
However, recent experimental findings \cite{johnson04} indicate that
interplay between Fano resonances and electron-electron interactions
may lead to interesting new effects.

Recently, there have been some attempts to understand the intermediate
coupling regime. Some of these efforts \cite{silvestrov00, berkovits05}
were confined to the case were only one dot level is coupled to the
leads, so that no interference can occur; while others
\cite{berkovits04, sindel05, gefen05, meden05, konik06, karrasch06}
discussed only a limited range of the parameter
space, (and thus did not mention, e.g., the possibility of
discontinuities for finite width levels), or considered solely the
question of the transmission phase \cite{golosov06}, not emphasizing
the behavior of the dot's population and its conductance. In this
paper we try to address those unexplored questions.

We discuss both the linear electric conductance, which
is the most easily accessible quantity experimentally, and the occupation
of the dot, which can be probed by, e.g., coupling it electrostatically
to a quantum point-contact. In fact, our model of a dot with several levels
can also describe a system of single level dots connected in parallel,
so that the occupation of each level can be measured separately.

After introducing out model and calculation methods in
Sec.~\ref{sec:model}, we briefly examine the non-interacting Fano
resonance in Sec.~\ref{sec:noninter}, and show that the observations by
Johnson \textit{et al.} \cite{johnson04}, interpreted by them as
interacting Fano resonances, can be explained as the result of
interference between one wide level and
many narrow levels in a non-interacting system.

In Sec.~\ref{sec:inter} we move to include interactions, which are
treated in a
Hartree-Fock approximation. We show that many new effects occur.
The interactions may lead to a non-monotonous filling of each
level as a function of an applied gate voltage. The dependence
on the gate voltage may be discontinuous, even when \emph{all}
the dot's levels have finite widths. In many cases both
continuous and discontinuous non-monotonicity can occur in the
same case. We find how the behavior of the population affects
the conductance. Interference effects between the dot levels play
a very important role in determining which type of behavior
will occur in a given sample.
We also discuss the effects of finite temperatures on the results,
which is essential for analyzing experimental data.

We conclude by reviewing our main findings in Sec.~\ref{sec:conclude}.

\section{\label{sec:model} Model and methods of calculation}

We consider the following model Hamiltonian, describing spinless
electrons (experimentally realizable by applying a strong in-plane
magnetic field) moving in a system composed of a (possibly
interacting) dot and non-interacting leads:
\begin{equation}
{\hat H} = {\hat H}_{D} + {\hat H}_{L} + {\hat H}_{R} + {\hat H}_{T}.
\end{equation}
This Hamiltonian is composed of three parts:
\begin {enumerate}
\item The quantum-dot Hamiltonian:
\begin{equation} \label{eqn:hd}
{\hat H}_{D} =
\sum_{i} \epsilon_{i,v} {\hat a}^{\dagger}_{i} {\hat a}_{i} +
\frac{U}{2} \sum_{i \ne j} {\hat a}^{\dagger}_{i} {\hat a}_{i}
{\hat a}^{\dagger}_{j} {\hat a}_{j}.
\end{equation}
Here ${\hat a}_{i}$ and ${\hat a}^{\dagger}_{i}$ are creation and
annihilation operators, respectively, of an electron in the dot's $i$'th
level; ${\epsilon}_{i,v} = {\epsilon}_{i} - eV_{g}$ is the corresponding
single-particle energy, modified by an applied gate voltage $V_{g}$ ($e$
is the absolute value of the electronic charge);
and $U=e^2/C$ is the strength of interaction between electrons in the
dot, assumed to consist simply of a charging energy.
\item The Hamiltonian of lead $\ell$ ( $=L$ or $R$ for the left or right
lead, respectively):
\begin{equation}
{\hat H}_{\ell} = \sum_{k} \epsilon_{k,\ell} {\hat c}^{\dagger}_{k,\ell}
{\hat c}_{k,\ell},
\end{equation}
where ${\hat c}^{\dagger}_{k,\ell}, {\hat c}_{k,\ell}$ are creation and
annihilation operators of an electron in the $\ell$'th leads $k$'th mode,
$\epsilon_{k,\ell}$ the corresponding single-particle energy.
In the following we assume that each
lead is a band of width $2D$, (much larger than any other energy scale
in the system), and constant density of states.
\item The tunneling Hamiltonian:
\begin{equation}
{\hat H}_{T} = \sum_{i,k,\ell}
\left( t^{i}_{k,\ell} {\hat a}^{\dagger}_{i} {\hat c}_{k,\ell}
+ H.C. \right),
\end{equation}
where the tunneling matrix elements $t^{i}_{k,\ell}$ are assumed
real (i.e., there is no applied out-of-plane magnetic field), and
independent of $k$. It is also assumed that
$\left| t^{i}_{L} \right| = \left| t^{i}_{R} \right|$.
\end{enumerate}
For $U=0$ (the non-interacting case), the Hamiltonian can be exactly
solved. The matrix elements (in the dot states space) of the inverse
retarded (advanced) Green function for the dot states is given by:
\begin{equation} \label{eqn:green}
{ \left( {G(\epsilon)}^{r,a} \right) }^{-1}_{i,j} =
\epsilon -  {\epsilon}_{i,v} {\delta}_{i,j} \pm
\frac{i}{2} \sum_{\ell=R,L} {\Gamma^\ell_{i,j}},
\end{equation}
where $\Gamma^\ell_{i,j}$, the matrix elements of the width of the
dot's levels due to their coupling with lead $\ell$, are given by:
\begin{equation}
{\Gamma}^{\ell}_{i,j} = 2 \pi { \left( t^{i}_{\ell} \right) }^{*}
t^{j}_{\ell} {\rho}_{\ell},
\end{equation}
${\rho}_{\ell}$ being the density of states in the $\ell$'th lead.
Thus, ${\left( \Gamma^\ell_{i,j} \right)}^{2} =
\Gamma^\ell_{i,i} \Gamma^\ell_{j,j}$. In addition, by the above
assumptions on the tunneling matrix elements,
$\left| \Gamma^L_{i,j} \right| = \left| \Gamma^R_{i,j} \right|$.
Thus, one is free to choose the diagonal matrix elements of
$\Gamma^L$,
and the \emph{signs} of the off-diagonal elements of both
$\Gamma^L$ and $\Gamma^R$.
In the following we will denote the total width, $\Gamma_L + \Gamma_R$
by $\Gamma$, and its $i$'th diagonal matrix elements by $\Gamma_i$.

Using the dot's Green functions, one can find the following averages,
related to the average occupation of the dot's states at temperature $T$:
\begin{equation} \label{eqn:n}
\langle \hat{a}^{\dagger}_{i} \hat{a}_{j} \rangle =
- \frac{1}{\pi} \int_{-D}^{D} f(\epsilon) \Im \{
 G^{r}_{i,j}(\epsilon) \} d\epsilon,
\end{equation}
$f(\epsilon) = 1 / ( \exp[(\epsilon-\mu)/T] + 1 )$ being the
Fermi-Dirac distribution function with chemical potential $\mu$ and
temperature $T$, using units where Boltzmann's constant equals unity.
In particular, the average occupation of the dot's $i$'th level
is $ n_i = \langle \hat{a}^{\dagger}_i \hat{a}_i \rangle $.

We can also find the linear conductance, following Meir and Wingreen
\cite{meir92}:
\begin{equation}
g = \frac{e^2}{h} \int_{-D}^{D} [ - {f}^{\prime}(\epsilon) ]
\mbox{Tr} \left[{G}^{r}(\epsilon) {\Gamma}_{L} {G}^{a}(\epsilon)
{\Gamma}_{R} \right] d\epsilon.
\end{equation}

The interacting case ($U \ne 0$) is treated using the self-consistent
Hartree-Fock approximation. This amounts to replacing the dot
Hamiltonian~(\ref{eqn:hd}) by an effective single-particle
Hamiltonian, given by:
\begin{eqnarray}
\nonumber \hat{H}^{eff}_{D} & = &
\sum_{i} \left( \epsilon_{i,v} + U \sum_{i} {n}_{i} \right)
{\hat a}^{\dagger}_{i} {\hat a}_{i} 
- U \sum_{i,j} \langle \hat{a}^{\dagger}_{i} \hat{a}_{j} \rangle
\hat{a}^{\dagger}_{j} \hat{a}_{i} \\ 
& & -\frac{U}{2} \left( \sum_{i} {n}_{i} \right)^2
+\frac{U}{2} \sum_{i,j}
\langle \hat{a}^{\dagger}_{i} \hat{a}_{j} \rangle^2 .
\end{eqnarray}
The diagonal terms in this expression are the Hartree contribution,
while Fock correction leads to the off-diagonal terms.
Using Eqs.~(\ref{eqn:green}) and (\ref{eqn:n}), the problem is reduced
to solving a set of self-consistent equations for the averages
$ \langle \hat{a}^{\dagger}_{i} \hat{a}_{j} \rangle $.
One can show that the Fock terms vanish for any pair of levels $i,j$
for which $\Gamma^L_{i,j} = - \Gamma^R_{i,j}$.

In many cases there are several solutions to the Hartree-Fock equations,
corresponding to different dot states being almost full or almost
empty. In those cases one should average over the solutions, giving
each a probability factor proportional to $\exp(-\Omega/T)$, where
$\Omega$ is the grand canonical free energy:
\begin{equation}
\Omega =  \frac{T}{\pi}
\int_{-D}^{D} \ln \left[ 1 + e^{ \frac{\mu-\epsilon}{T} } \right]
 \Im \{ \mbox{Tr} \left[ G^{r}(\epsilon) \right] \} d\epsilon.
\end{equation}
In particular, at zero temperature only the solution with the lowest
chemical potential should be retained.

We mention in passing that the above integrals for the free energy,
occupations and conductance can be expressed in terms of the logarithm
of the gamma function of a complex argument and its first two derivatives
(the digamma and trigamma functions) \cite{abramsteg}, respectively,
to facilitate faster computation.

\section {\label{sec:noninter} Interference effects in non-interacting dots}

\begin{figure*}
\includegraphics[width=12cm,height=12cm]{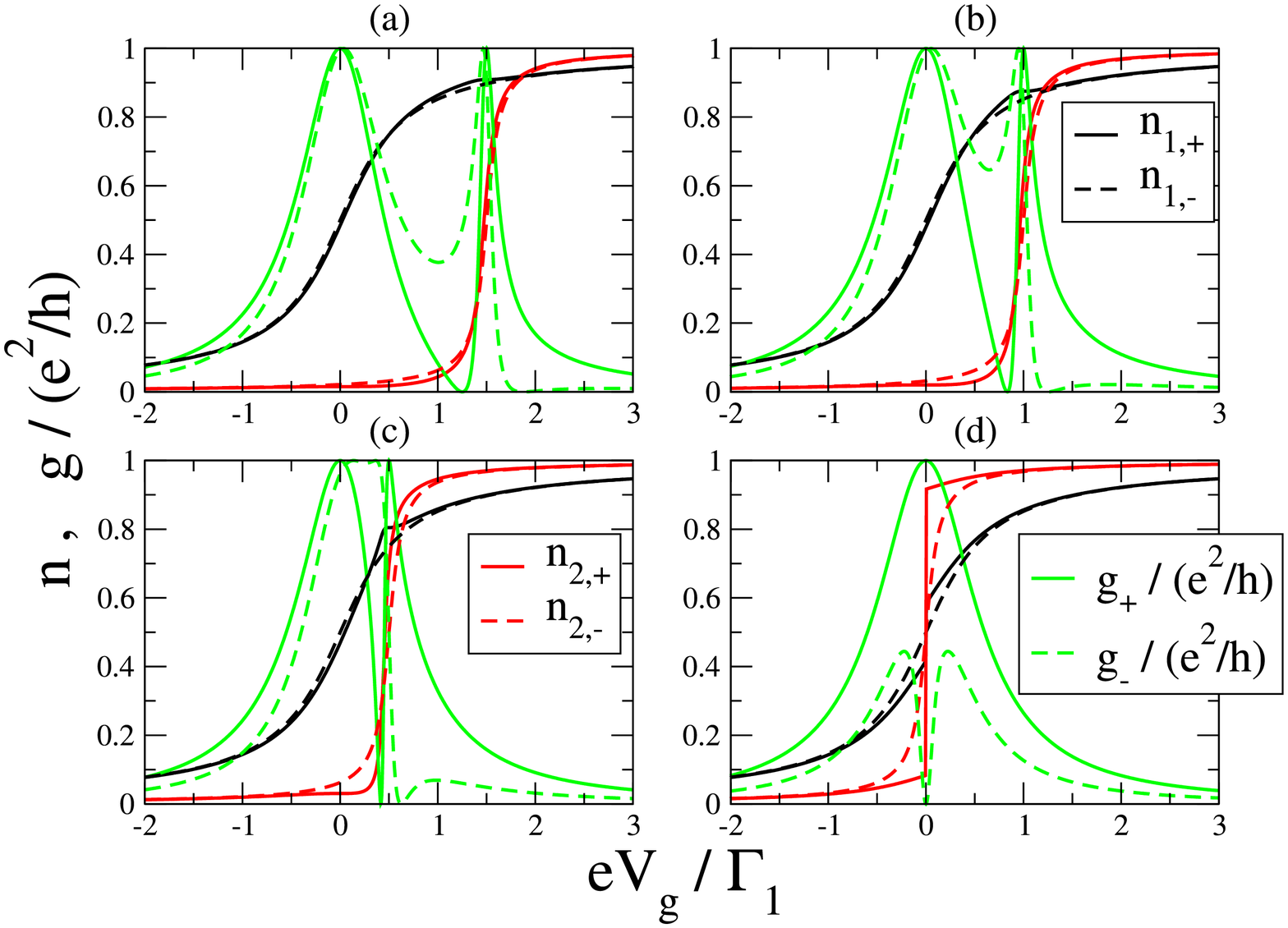}
\caption{\label{fig:noninter} Level occupations and conductance of
a two-level non-interacting dot at zero temperature. In all the
graphs, $\epsilon_1/\Gamma_1=0.0$, $\Gamma_2/\Gamma_1=0.2$, while
$\epsilon_2$ varies:
(a) $\epsilon_2/\Gamma_1=1.5$;
(b) $\epsilon_2/\Gamma_1=1.0$;
(c) $\epsilon_2/\Gamma_1=0.5$;
(d) $\epsilon_2/\Gamma_1=0.0$.}
\end{figure*}

\begin{figure}
\includegraphics[width=8cm,height=8cm]{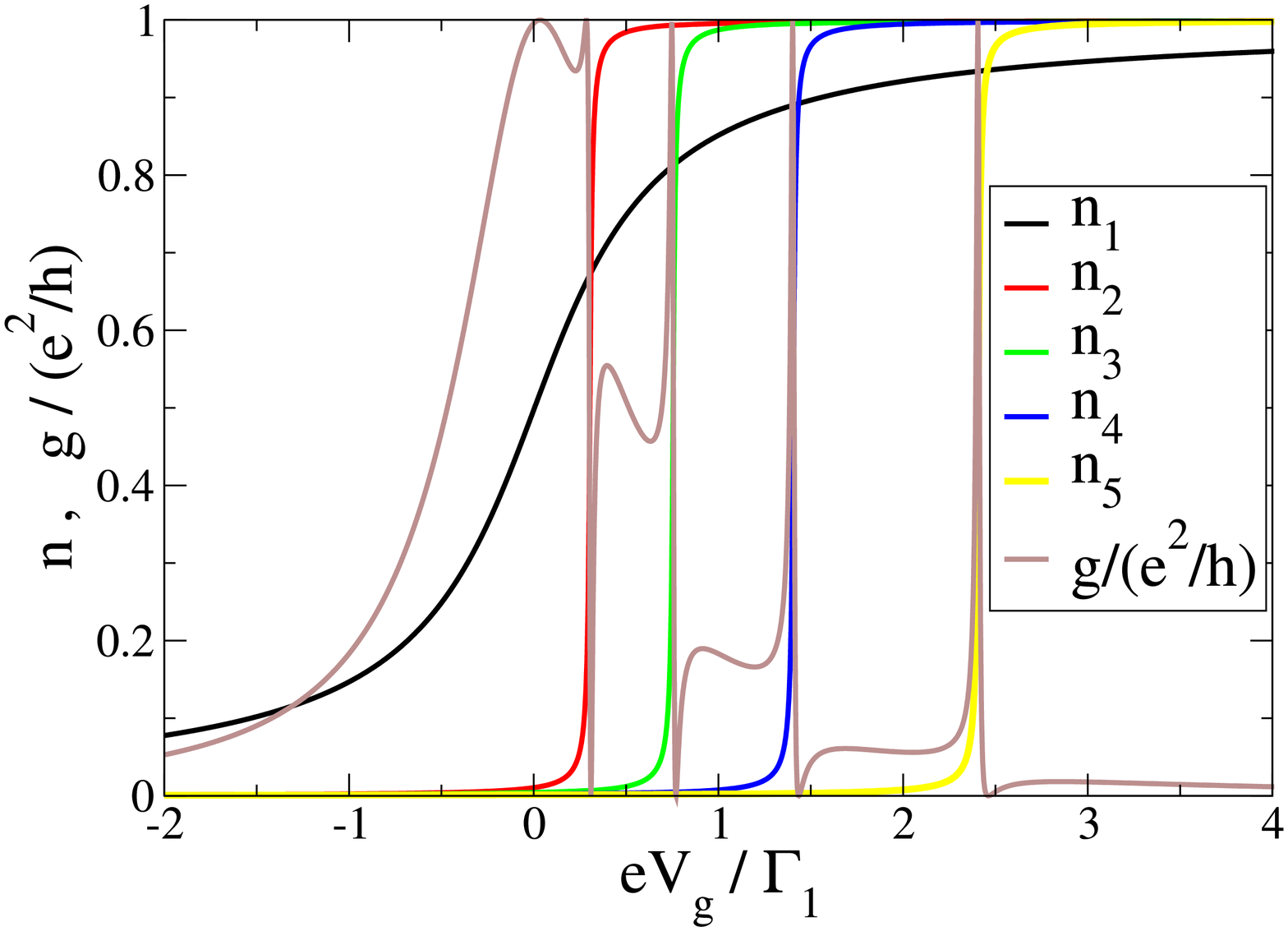}
\caption{\label{fig:marcus} Level occupations and conductance of a
non-interacting dot with a single broad level and four narrow levels,
at zero temperature. The wide level is
coupled to one of the leads with a sign opposite to that of the narrow
levels. The levels are at
$\epsilon_i/\Gamma_1 = \{0.0, 0.3, 0.75, 1.4, 2.4\}$, with widths
$\Gamma_i/\Gamma_1 = \{1.0 ,0.02, 0.02, 0.02, 0.02\}$, for i=1...5.
The general shape of the conductance function
resembles Fig.~1 of Ref.~\onlinecite{johnson04}. }
\end{figure}

We first consider the non-interacting case. Here, for a two-level dot,
there are two possibilities, depending on the relative signs of the
matrix elements of the widths $\Gamma_{R,L}$ between the two states.
These will be termed the plus (minus) configuration for positive
(negative) sign of $\Gamma^{R}_{1,2}/\Gamma^{L}_{1,2}$.
If we make a transformation from the lead operators $\hat{c}_{k,R},
\hat{c}_{k,L}$ to the combinations $\hat{c}_{k,\pm} = ( \hat{c}_{k,R}
\pm \hat{c}_{k,R} ) / \sqrt{2}$, we find that in the plus case the
two dot states are connected to the $\hat{c}_{k,+}$ states,
whereas in the minus
case one dot state is connected to the $\hat{c}_{k,+}$ states,
while the other is connected with the $\hat{c}_{k,-}$ states.
Thus, in the plus case the two dot states are effectively
coupled to a single lead, while in the minus case each dot
state is connected to a different effective lead \cite{sindel05}.

The local density of states of each level, $\rho_i$, is given by
($i^\prime$ is the index of the other level):
\begin{subequations}
\begin{eqnarray}
\nonumber \rho^{+}_i(\epsilon) &=& \frac{1}{\pi}
\frac{ \frac{\Gamma_i}{2} ( \epsilon - \epsilon_{i^\prime,v} )^2 }
{ ( \epsilon - \epsilon_{i,v} )^2
( \epsilon - \epsilon_{i^\prime,v} )^2 +
\left( \frac{ \Gamma_i + \Gamma_{i^\prime} } {2} \right)^2
(\epsilon - \epsilon_{+,v} )^2 }, \\ & & \\
\rho^{-}_i(\epsilon) &=& \frac{1}{\pi} \frac{ \frac{\Gamma_i}{2} }
{ ( \epsilon - \epsilon_{i,v} )^2 +
\left( \frac{\Gamma_i}{2} \right)^2 },
\end{eqnarray}
\end{subequations}
where $\epsilon_{\pm,v} =
( \Gamma_1 \epsilon_{2,v} \pm \Gamma_2 \epsilon_{1,v} )
/ ( \Gamma_1 \pm \Gamma_2 )$.
As one can see, in the minus case, the density of states of
each of the dot's levels is unaffected by the other level, because
they are effectively decoupled, as was explained above. Thus, their
populations follow the usual
$n_i=1/2+\tan^{-1}[(\mu-\epsilon_{i,v})/(\Gamma_i/2)]/\pi$ low.
However, in the plus case, the levels interfere.
As a result, the local density of states of each level goes to zero at
the position of the other level. Thus, as the two levels approach each
other, the density of states for both levels develops a sharp peak,
going from zero to its maximum in a gate voltage distance which goes as
$~|\epsilon_1-\epsilon_2|$, instead of the widths $\Gamma_i$. This
causes the populations $n_i$ to vary fast for $V_g$ between the two
level energies. When the level energies exactly coincide, this sharp
feature becomes a delta function peak in the density of states, or a
discontinuous jump in the level population as a function of the gate
voltage. Indeed, in this latter case, due to the degeneracy of the
dot levels, one can transform to a basis of the states where one
level is totally decoupled from the leads\cite{konig98, berkovits04}.

The conductance in each case is given by:
\begin{subequations} \label{eqn:gpm}
\begin{eqnarray}
\nonumber g^{+} &=& \frac{e^2}{h} \frac {
\left( \frac{ \Gamma_1 + \Gamma_2 } {2} \right)^2
(\mu - \epsilon_{+,v} )^2 }
{ ( \mu - \epsilon_{1,v} )^2 ( \mu - \epsilon_{2,v} )^2 +
\left( \frac{ \Gamma_1 + \Gamma_2 } {2} \right)^2
(\mu - \epsilon_{+,v} )^2 } , \\ & & \\
\nonumber g^{-} &=& \frac{e^2}{h} \frac {
\left( \frac{ \Gamma_1 - \Gamma_2 } {2} \right)^2
( \mu - \epsilon_{-,v} )^2 }
{ \left[ (\mu - \epsilon_{1,v})^2 +
\left( \frac{\Gamma_1}{2} \right)^2 \right]
\left[ (\mu - \epsilon_{2,v})^2 +
\left( \frac{\Gamma_2}{2} \right)^2 \right] } . \\ & &
\end{eqnarray}
\end{subequations}
Since conductance occurs by transmission of electrons between the left and
right leads, and not between the $\hat{c}_{k,\pm}$ combinations, we find
interference effects in the conductance in \emph{both} cases.
Thus, in the plus case, the conductance reaches its maximal possible
value (of $e^2/h$) when the $\mu$ equals one of the dot's levels
($\epsilon_1$ or $\epsilon_2$), and goes to zero for $\mu = \epsilon_{+}$,
i.e., between the conductance peeks.
In the minus case, the conductance reaches its maximal value for
$\mu$ in the vicinity of (but slightly different from) the dot's levels.
It goes to zero at $\mu = \epsilon_{-}$, which lies outside the peaks.
When $\epsilon_1$ and $\epsilon_2$ are sufficiently different, the
conductance reaches its maximal possible value of $e^2/h$ at the peaks.
When the dot's levels are too close the two peaks near the level energies
become smaller, and eventually
merge into a single peak. Finally, For $\epsilon_1 = \epsilon_2$ and
$\Gamma_1 = \Gamma_2$, there is a complete destructive interference
between the two levels, and $g=0$ for all $\mu$ values.

\begin{figure}
\includegraphics[width=8cm,height=8cm]{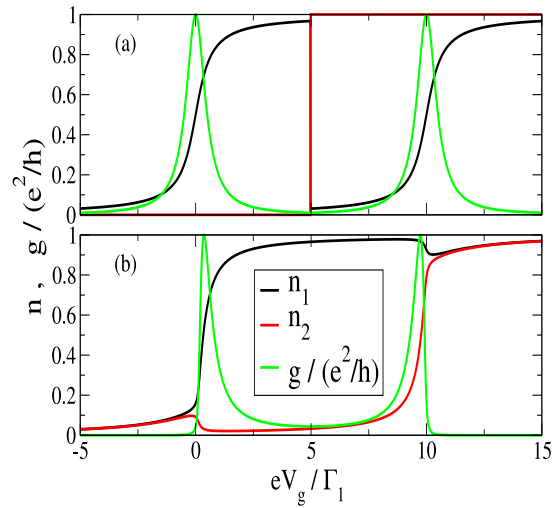}
\caption{\label{fig:nonmon} The two basic phenomena of non-monotonic
charging.
The graph shows level occupations and conductance of a two-level
interacting dot at zero temperature. The two levels are connected
in the minus configuration.
In both graphs, $\epsilon_1/\Gamma_1=0.0$, $U/\Gamma_1=10.0$.
(a) effect \ref{itm:mech_a}: $\epsilon_2/\Gamma_1=0.0$,
$\Gamma_2/\Gamma_1=0.0$;
(b) effect \ref{itm:mech_b}: $\epsilon_2/\Gamma_1=0.1$,
$\Gamma_2/\Gamma_1=1.0$.
Consult the text for further explanation.}
\end{figure}

All the above results are exemplified in Fig.~\ref{fig:noninter}. Here,
and in the following, we set $\mu=0$, and vary the gate voltage $V_{g}$.
One can see that the interference effects result in an asymmetric shape
of the conductance peaks, usually referred to as ``Fano Resonances''
\cite{gores00, kobayashi02, clerk01, alhassid03, johnson04}.

We note that if the couplings to the left and right leads were different
in magnitude, we would get the same qualitative picture, but the
conductance would have reached values lower than $e^2/h$ even at the
peaks in all cases.

One can show the above results for the conductance obey the the
relation $g^{\pm}=e^2/h \times \sin^2[\pi(n_1 \pm n_2)]$ at zero
temperature; this relation is also obeyed in the interacting case, in our
Hartree-Fock approximation. The validity of this equation is, however,
much wider, since it is required by the Friedel sum rule
\cite{datta97, kashcheyevs06}.

\begin{figure*}
\includegraphics[width=13cm,height=13cm]{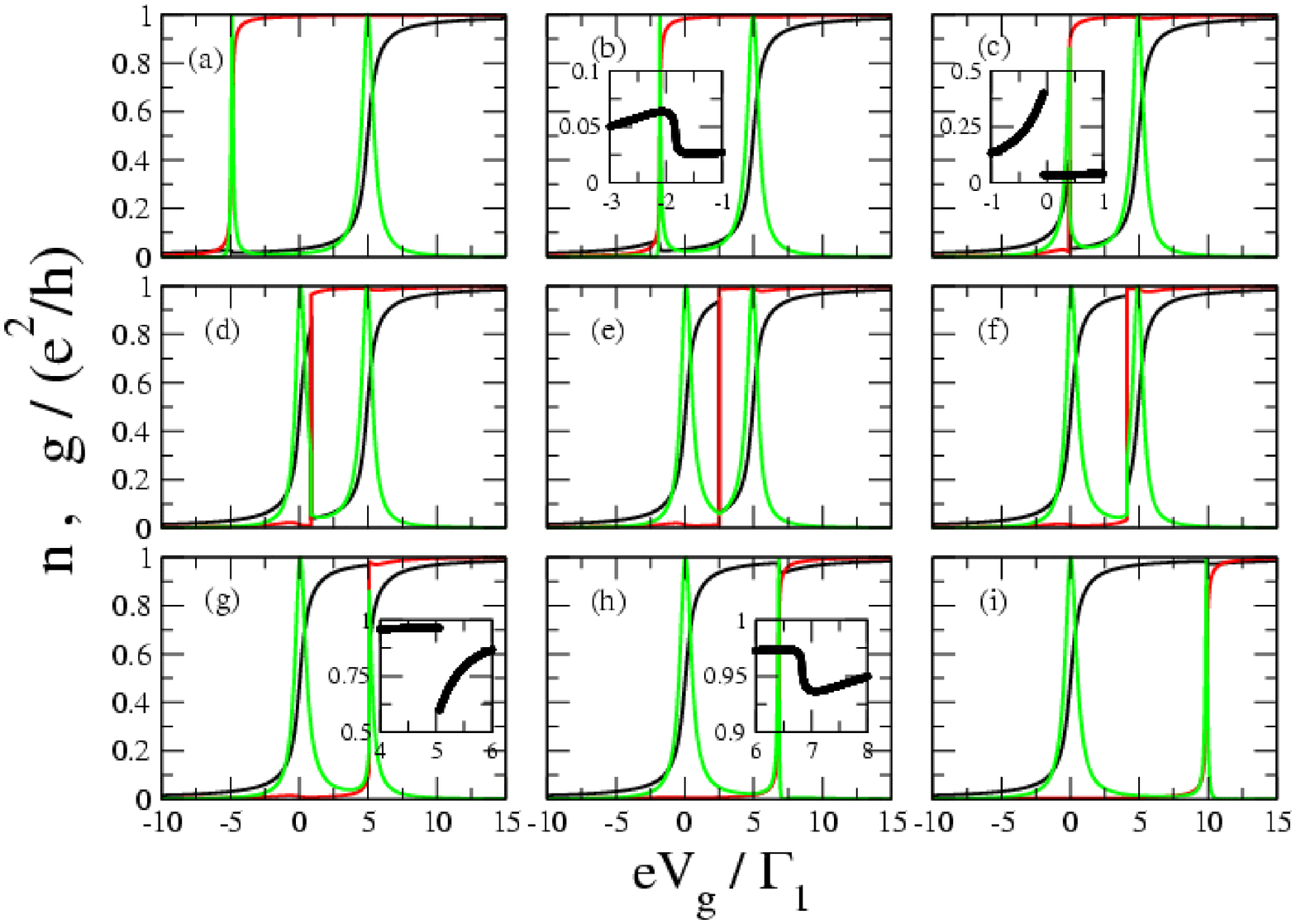}
\caption{\label{fig:mww} Level occupations (level 1 - black line; level
2 - red line) and conductance (green line) of a two-level interacting dot
at zero temperature. The two levels are connected in the minus
configuration.
In all the graphs, $\epsilon_1/\Gamma_1=0.0$,
$\Gamma_2/\Gamma_1=0.2$ and $U/\Gamma_1=5.0$ while
$\epsilon_2$ varies:
(a) $\epsilon_2/\Gamma_1=-5.0$;
(b) $\epsilon_2/\Gamma_1=-2.0$;
(c) $\epsilon_2/\Gamma_1=-0.5$;
(d) $\epsilon_2/\Gamma_1=-0.2$;
(e) $\epsilon_2/\Gamma_1=0.0$;
(f) $\epsilon_2/\Gamma_1=0.2$;
(g) $\epsilon_2/\Gamma_1=0.5$;
(h) $\epsilon_2/\Gamma_1=2.0$;
(i) $\epsilon_2/\Gamma_1=5.0$.
The insets to panels (b), (c), (g) and (h) show $n_1$ in black circles
in the region of fast variation, showing clearly that in cases (b) and (h)
the variation is continuous [like cases (a) and (i)], while in cases
(c) and (g) it is discontinuous [like cases (d)--(f)]
(to an accuracy in $V_g/\Gamma_1$ better than ${10}^{-3}$).}
\end{figure*}

As one can see, in the minus case, we get three conductance peaks
from only two dot levels. As we show in Fig.~\ref{fig:marcus}, this
can be extended to the case of many narrow levels and a single wide level,
where the latter's coupling to one of the the leads has an opposite sign
(in the sense discussed above) to that of the narrow levels. In this way
we get two peaks in the conductance for each narrow levels. The conductance
curve is very similar to some of the experimental results of Johnson
\textit{et el.} \cite{johnson04}. This shows that interference effects
alone can qualitatively explain
the experiments, without the need to resort to interaction effects, as
was done by the above mentioned authors.

\section {\label{sec:inter} Interaction effects}

We now turn on the interactions. We will focus on two-level systems
from now on. The simplest effect of the interactions is that a
filled level pushes unfilled ones toward higher energies by the
Hartree term.
However, interactions can lead to much more interesting phenomena, such
as non-monotonous population of the levels. In the two limiting
cases in terms of the ratio between the widths of the two levels (i.e.,
one of the levels has zero width, or has the same width as the other level
respectively), one finds either of the two most basic effects:

\begin{figure*}
\includegraphics[width=13cm,height=13cm]{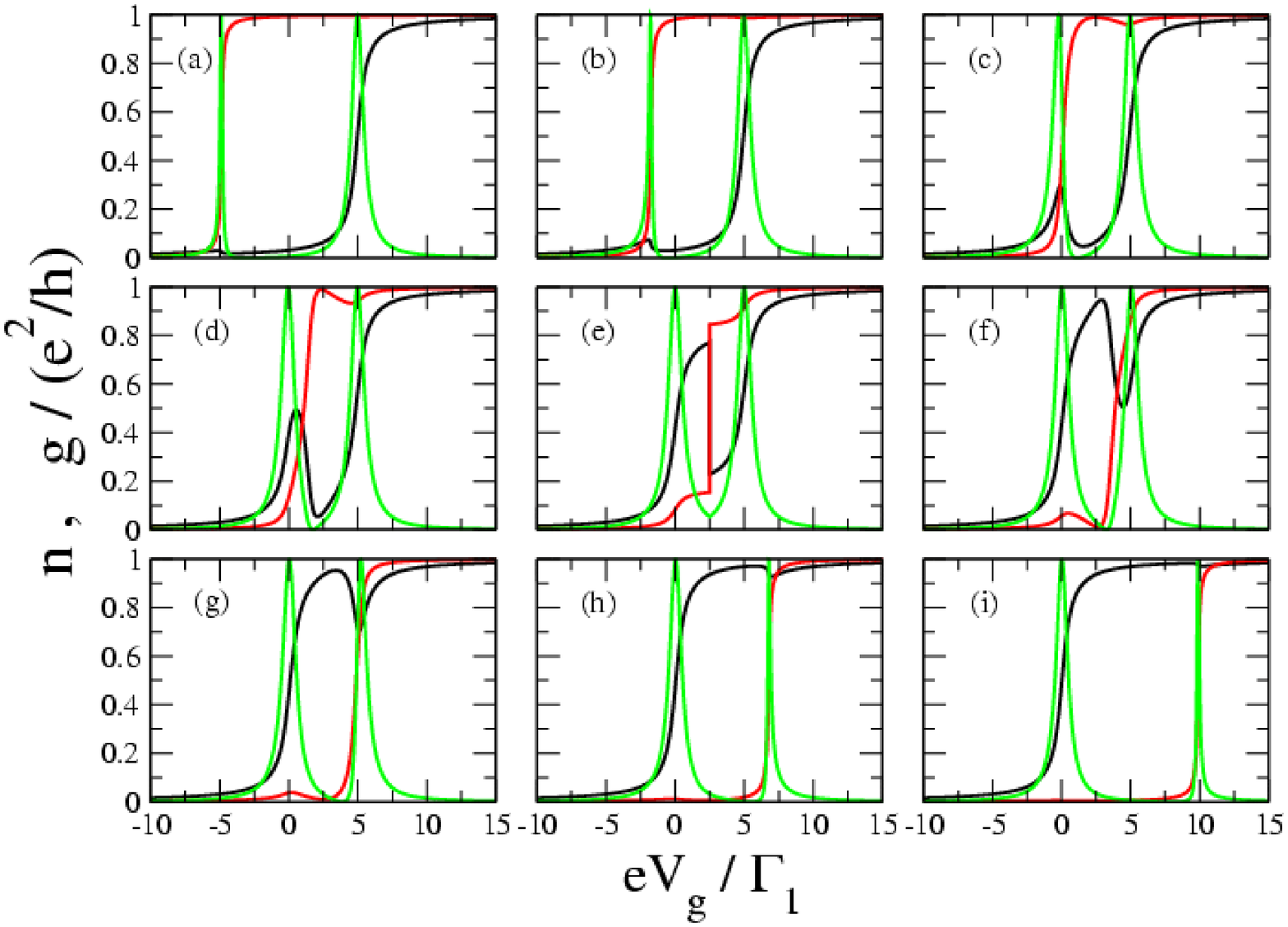}
\caption{\label{fig:pww} Level occupations (level 1 - black line; level
2 - red line) and conductance (green line) of a two-level interacting dot
at zero temperature. The two levels are connected in the plus
configuration.
In all the graphs, $\epsilon_1/\Gamma_1=0.0$,
$\Gamma_2/\Gamma_1=0.2$ and $U/\Gamma_1=5.0$ while
$\epsilon_2$ varies:
(a) $\epsilon_2/\Gamma_1=-5.0$;
(b) $\epsilon_2/\Gamma_1=-2.0$;
(c) $\epsilon_2/\Gamma_1=-0.5$;
(d) $\epsilon_2/\Gamma_1=-0.2$;
(e) $\epsilon_2/\Gamma_1=0.0$;
(f) $\epsilon_2/\Gamma_1=0.2$;
(g) $\epsilon_2/\Gamma_1=0.5$;
(h) $\epsilon_2/\Gamma_1=2.0$;
(i) $\epsilon_2/\Gamma_1=5.0$.
In this figure there is no discontinuity in any of the cases.}
\end{figure*}

\begin{figure*}
\includegraphics[width=13cm,height=13cm]{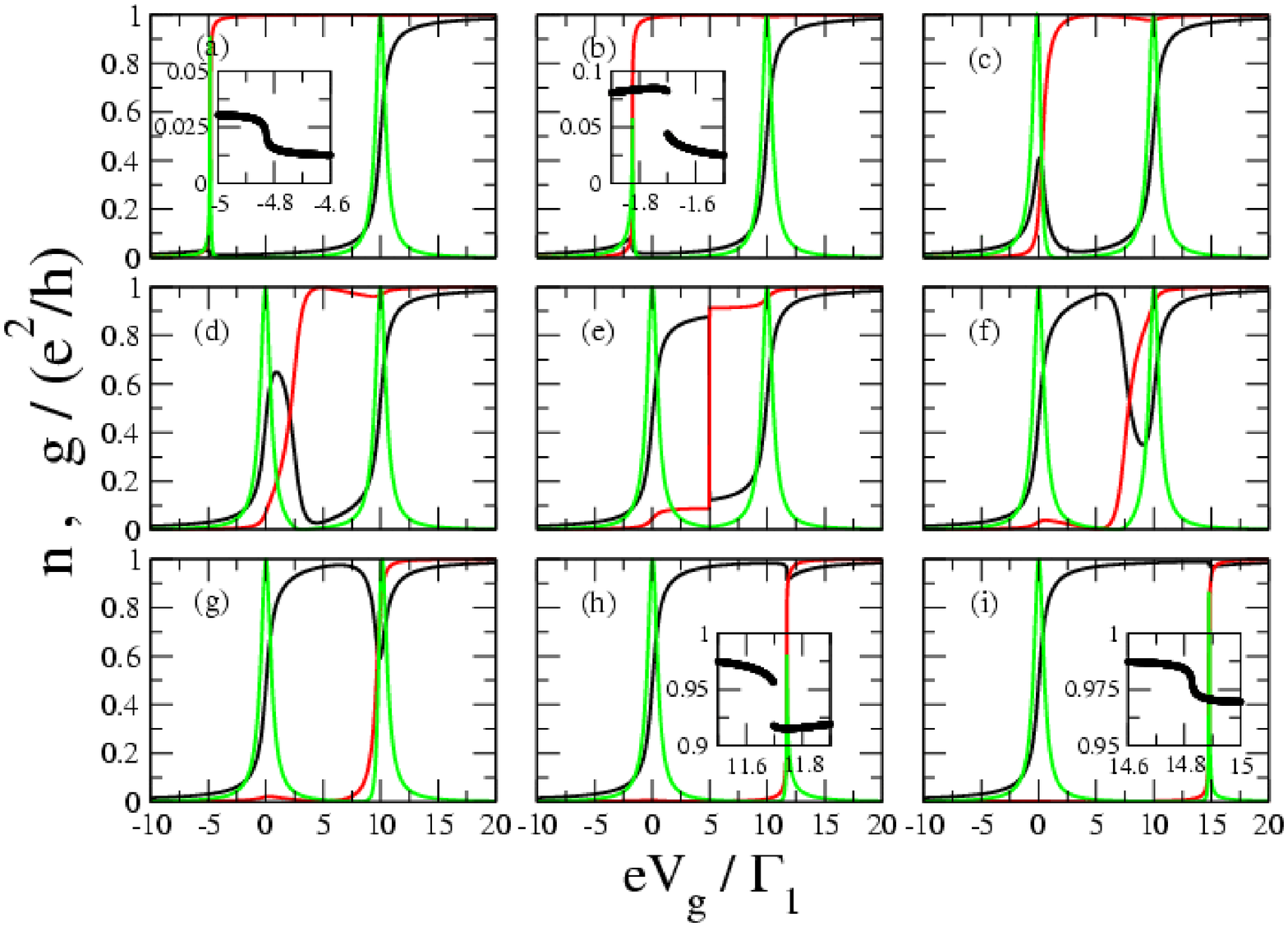}
\caption{\label{fig:pns} Level occupations (level 1 - black line; level
2 - red line) and conductance (green line) of a two-level interacting dot
at zero temperature. The two levels are connected in the plus
configuration.
In all the graphs, $\epsilon_1/\Gamma_1=0.0$,
$\Gamma_2/\Gamma_1=0.1$ and $U/\Gamma_1=10.0$ while
$\epsilon_2$ varies:
(a) $\epsilon_2/\Gamma_1=-5.0$;
(b) $\epsilon_2/\Gamma_1=-2.0$;
(c) $\epsilon_2/\Gamma_1=-0.5$;
(d) $\epsilon_2/\Gamma_1=-0.2$;
(e) $\epsilon_2/\Gamma_1=0.0$;
(f) $\epsilon_2/\Gamma_1=0.2$;
(g) $\epsilon_2/\Gamma_1=0.5$;
(h) $\epsilon_2/\Gamma_1=2.0$;
(i) $\epsilon_2/\Gamma_1=5.0$.
The insets to panels (a), (b), (h) and (i) show $n_1$ in black circles
in the region of fast variation. One can see the variation is continuous in
all cases except (b) and (h), where it is discontinuous
(to accuracy in $V_g/\Gamma_1$ better than ${10}^{-3}$). }
\end{figure*}

\begin{enumerate}
\item \label{itm:mech_a} As was first noted by Silvestrov and Imry
\cite{silvestrov00}, when
one of the levels is completely decoupled from the lead (i.e., has zero
width), it is either completely filled or completely empty at zero
temperature. As the gate voltage is swept from low to high values,
the wider level is first filled (if its energy is not too much higher
than that of the narrow level),
and pushes up the energy of the narrow level. Indeed, for low
gate voltage values, filling the wider level gives lower total
kinetic energy than filling the narrow one. However, At some point,
for higher gate voltage values, the kinetic energy considerations make
it advantageous to fill the narrow level instead , thus moving the
wider one toward higher energy values, and reducing its population.
Thus, at that point there is a sharp jump in the occupation of both
levels, and in the conductance. Also, the two conductance peaks
result from filling the same level twice, and thus have equal
lineshapes.
\item \label{itm:mech_b} For levels of comparable widths, no
discontinuity occurs. As the
gate voltage is swept, the two levels start becoming populated.
However, if one level has a lower bare energy, or is wider, this
level gets populated faster, and repulses the electrons out of the other
level, thus increasing even further its own population.
When the gate voltage becomes higher, the process is reversed: the
less-populated level starts to be occupied again, and in turn reduces
the population of the more-populated level. This effect causes the
conductance peaks to be asymmetric. This effect is most
pronounced when the level widths are in fact equal.
\end{enumerate}

We will begin by examining the case of levels in the minus configuration.
This is the simpler case, since, as we have seen in the non-interacting
case, the interference effects do not affect the dot's population. To
start with, we give examples to the two limiting cases discussed above in
Fig.~\ref{fig:nonmon}. One can observe that, in contrast with the
non-interacting case discussed in the previous section, the conductance
does not go to zero in both cases - in the first, one level is completely
decoupled from the rest of the system (except for the interaction), so
no interference can occur even in the conductance. In the second case, the
levels have equal widths, and $\epsilon_{-,v}$ goes to infinity (See
Eq.~(\ref{eqn:gpm})).

In intermediate cases between those two limits, all kinds of combinations
of these two phenomena can occur. Some typical cases are shown in
Fig.~\ref{fig:mww}. In all those cases the interaction is much stronger
than the level widths, and the width of level 2 is much smaller than the
width of level 1, but the distance between the two levels varies.

The most interesting feature of the results is that the phenomena of
discontinuity of the dot's properties (level populations and conductance)
as a function of $V_g$ is not
restricted to the case of a level of zero width shown above, but can
appear even when the widths of both levels are finite. As one can see in
the figure, the discontinuity occurs when the energies of the levels
are sufficiently close [cases (c)--(g)]; otherwise, the variation is
continuous. By varying the other parameters we have found that for any
given value of level separation (including zero), this effect occurs
if the interaction is strong enough, and the ratio between the width
of the narrow level and the width of the wide level is small enough
--- The limiting values are less restrictive for close levels,
and more demanding for far-away ones.

In addition to this effect, the second, continuous effect of
non-monotonous level filling can also be observed in cases (c)--(g).
However, it is quite weak here, since the level widths are far from
being equal.

Considering the conductance, we have seen in the non-interacting case
that it goes to zero at a value of $V_g$ on the narrower peak side
away from the wider peak. This effect is manifested
only when there is no discontinuity, and results in an
asymmetry of the narrower peak. In cases (c)--(g), the
discontinuity skips over this point, but leaves the narrow
peak asymmetric. In addition, since the wider level is filled in both
the conductance peaks, it makes the narrow peak wider in comparison
with the non-interacting case (Although, in contrast with the case
of Fig.~\ref{fig:nonmon}(b), the narrow level has a finite width, and
the two conductance peaks have \emph{different} lineshapes).

The situation is, however, quite different in the plus configuration.
Here, there is a strong interference effect on the populations of the two
levels. This is due to both the off diagonal element of the total widths
matrix $\Gamma$, (which create imaginary off diagonal elements in the
dot's inverse Green function),
and the Fock term (which create real off diagonal elements in the
dot's inverse Green function).
Both these terms are absent in the minus configuration discussed above.

\begin{figure*}
\includegraphics[width=13cm,height=10cm]{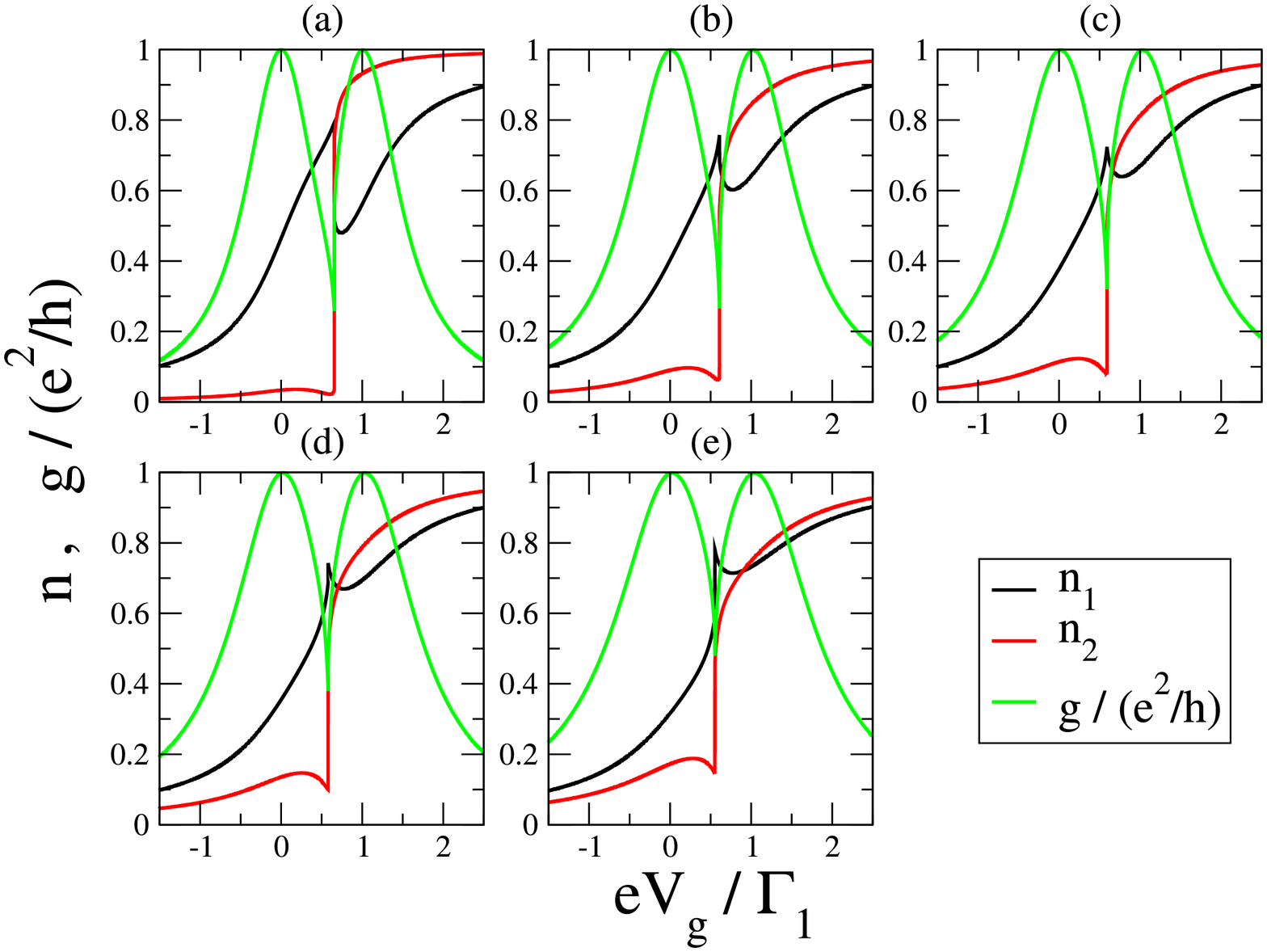}
\caption{\label{fig:pvc1} Level occupations and conductance of a
two-level interacting dot at zero temperature. The two levels are
connected in the \emph{plus} configuration.
In all the graphs, $\epsilon_1/\Gamma_1=0.0$,
$\epsilon_2/\Gamma_1=0.1$ and $U/\Gamma_1=1.0$ while
$\Gamma_2$ varies:
(a) $\Gamma_2/\Gamma_1=0.1$;
(b) $\Gamma_2/\Gamma_1=0.3$;
(c) $\Gamma_2/\Gamma_1=0.4$;
(d) $\Gamma_2/\Gamma_1=0.5$;
(e) $\Gamma_2/\Gamma_1=0.7$.
In cases (a) and (b) $n_1$ increases at the discontinuity, in case
(c) it does not change, while in cases (d) and (e) it decreases at the
discontinuity. In all the cases, $n_2$ increases at the discontinuity.}
\end{figure*}

Fig.~\ref{fig:pww} shows the results in the plus configuration,
where all parameters values are the same as in Fig.~\ref{fig:mww}.
Because of the interference effects on the populations of the two
levels, there are no discontinuities here, except in the case of
exactly degenerate levels [case (e)], where we observe the
\emph{non-interacting discontinuity}, which occurs even for free electrons
(as we have seen in Fig.~\ref{fig:noninter}(d) --- This point will be
discussed more fully later on).

Instead of discontinuities we observe in Fig.~\ref{fig:pww} a
continuous version of effect~\ref{itm:mech_a}, where the narrower level
depopulates the wider one as $V_g$ is increased (like, e.g., case
(d) for $V_g/\Gamma_1$ near zero), resulting in a broadening of the
of the narrow conductance peak. We also observe effect~\ref{itm:mech_b},
where the inverse occurs (such as, e.g., case (f) for $V_g/\Gamma_1$
near zero), contributing to the asymmetry of the wide peak.

We also note that here, in all the cases [except
the exactly degenerate case (e)] there is a value of $V_g$ between
the two conductance peaks where the conductance goes to zero, as for
the non-interacting case.

Nevertheless, we can obtain discontinuities even in the plus configuration.
This is exemplified in Fig.~\ref{fig:pns}, in which the parameters are
similar to Fig.~\ref{fig:pww}, but the interaction is stronger and the
ratio between the narrow level width and the wide level width is smaller.
As in the minus case, there is no discontinuity when the levels are too far
away [cases (a) and (i)]. In contrast with the minus case, interference 
effects prevent the discontinuities when the levels are too close [cases
(c)--(g), except the non-interacting discontinuity for the exactly
degenerate case (e)]. Only for the intermediate cases (in terms of energy
levels distance) does discontinuity occurs [as can be seen in the insets
to panels (b) and (h)].

In addition, continuous non-monotonicity of the level populations (and the
resulting broadening and asymmetry of the narrow peak) occurs
in cases (c), (d), (f) and (g), in a similar way to the corresponding cases
in Fig.~\ref{fig:pww}. Here too the conductance vanishes
for some value of $V_g$ between the two conductance peaks in all cases
[except (e)].
This happens even in the discontinuous cases [in contrast with the
situation in the minus configuration, Fig.~\ref{fig:mww}(c)--(g)].
The reason is the smallness of the discontinuity even when it occurs. 

\begin{figure*}
\includegraphics[width=10cm,height=10cm]{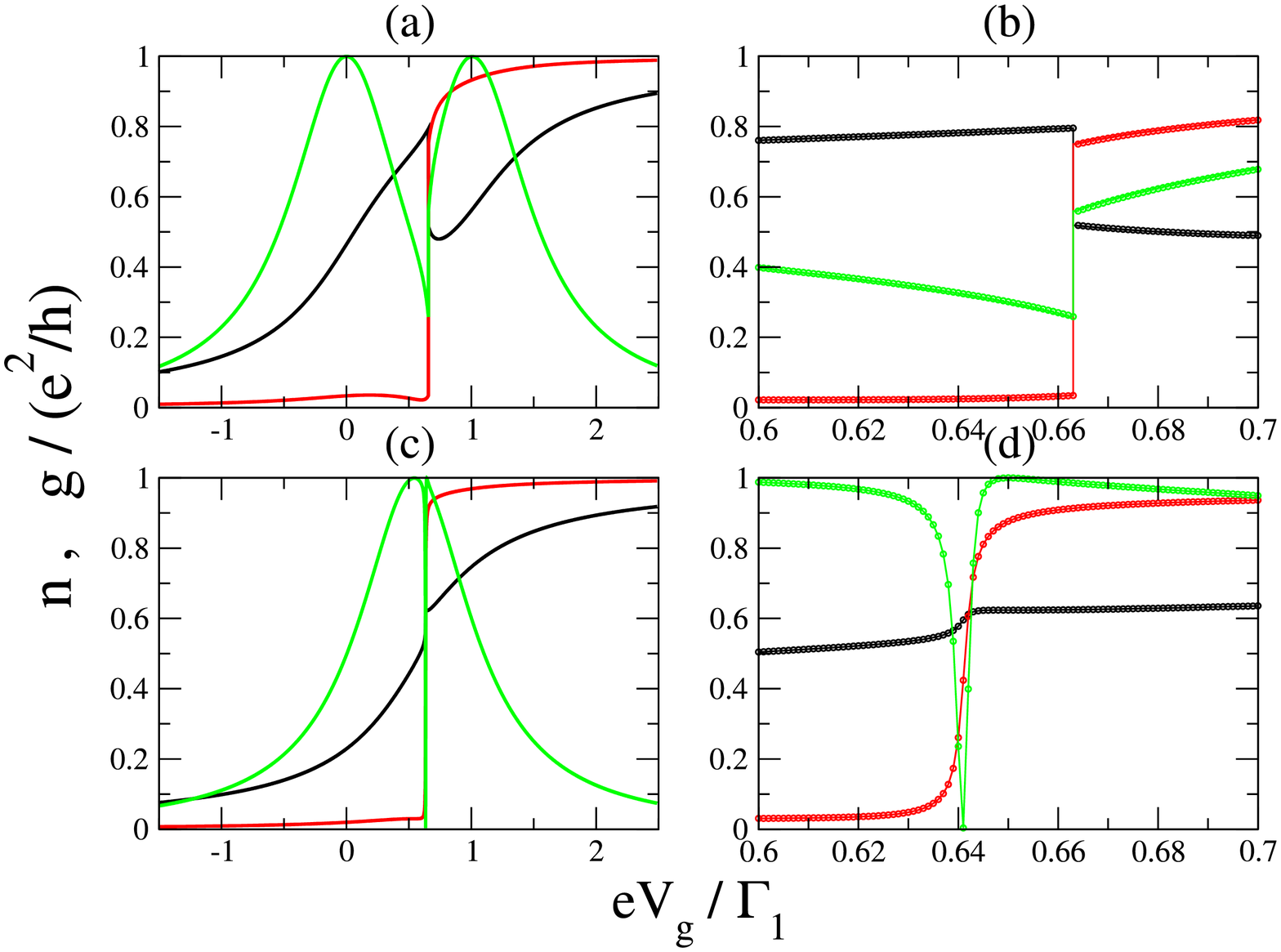}
\caption{\label{fig:pvc2} Level occupations (level 1 - black line; level
2 - red line) and conductance (green line) of a two-level dot
at zero temperature. The two levels are connected in the plus
configuration.
In all the graphs, $(\epsilon_2-\epsilon_1)/\Gamma_1=0.1$ and
$\Gamma_2/\Gamma_1=0.1$.
(a-b) $\epsilon_1/\Gamma_1=0.0$, $U/\Gamma_1=1.0$ - panel
(b) is a close-up on the sharp features of panel (a);
(c-d) $\epsilon_1/\Gamma_1=0.55$, $U/\Gamma_1=0.0$ - panel
(d) is a close-up on the sharp features of panel (c).
In the non-interacting case the variation is very sharp but continuous;
in the interacting case there is a discontinuity.}
\end{figure*}

In Sec.~\ref{sec:noninter} we have seen that for a non-interacting
dot in the plus configuration, when the two levels are almost
degenerate, they
can be treated as a linear combination of a wide and a very narrow
effective levels, almost decoupled for each other. The presence of the
narrow effective level causes a sharp variation in the population of the
dot's original levels as a function of $V_g$. In the limit of
exactly degenerate levels, the narrow level width become zero (it
becomes completely decoupled), so a discontinuity occurs
\cite{konig98, berkovits04}.

The results shown so far seem to imply that in the presence of
interaction this sharp variation disappears, except for the
case of coinciding levels, where the non-interacting discontinuity
is still observed. However, This is only true for strong
interactions, which generate a strong Fock interference term between
the levels. For weak interactions (too weak for discontinuities at
intermediate level separations to occur), the situation is quite
different. Here, because of the presence of the effective narrow
level, the interactions can turn the sharp but continuous variation
of the level populations in the non-interacting case into a
discontinuous one. The jump actually occurs for the effective
levels, increasing the population of the narrow effective level
by almost one, and reducing the population of the wider effective
level by a smaller amount (since the interaction is relatively
weak).

A typical situation is exhibited in Fig.~\ref{fig:pvc1}. Here the
first dot level width, the level distance and interaction are fixed,
but the second dot level width varies. When the second dot level is much
narrower than the wide one [cases (a) and (b)], the picture
is quite similar to the usual interacting discontinuity -- at the
discontinuity the population of the narrow dot level rises, while
that of the wide dot level falls. This is because the narrow dot level
is composed mainly of the narrow effective level, while
the wide dot level is composed mainly of the wide effective level.
When the narrow dot level width is comparable to the of the wider dot
level [cases (d) and (e)], the results resemble the
non-interacting discontinuity -- the occupations of both the
wider and the narrower dot levels rise at the discontinuity.
This happens since the wider dot level now has a larger share in the
narrow effective level, and, as was explained above, the increase
in the population of the narrow effective level is larger than the
(absolute value of the) decrease in the population of the wide
effective level. There is an intermediate value of the narrow dot
level width [case (c)] where only the narrower dot level population
jumps, while the wider dot level population is continuous, since
the effects of the wide and narrow effective levels on its
population exactly cancel.

Due to the large discontinuities in these cases, the conductance
does not vanish between the two peaks in any of the cases
considered. The result is two wide and overlapping conductance
peaks, with discontinuous features in the conductance valleys
between them.

To clarify the above points, in Fig.~\ref{fig:pvc2} we compare
the results with and without
interaction for parameter values corresponding to
Fig.~\ref{fig:pvc1}(a). It can be clearly seen that although
in the non-interacting case the variation of the physical
parameters (especially the conductance) with $V_g$ is very
fast, it is still continuous, and discontinuities can appear only
with the addition of interactions.

\begin{figure*}
\includegraphics[width=12cm,height=12cm]{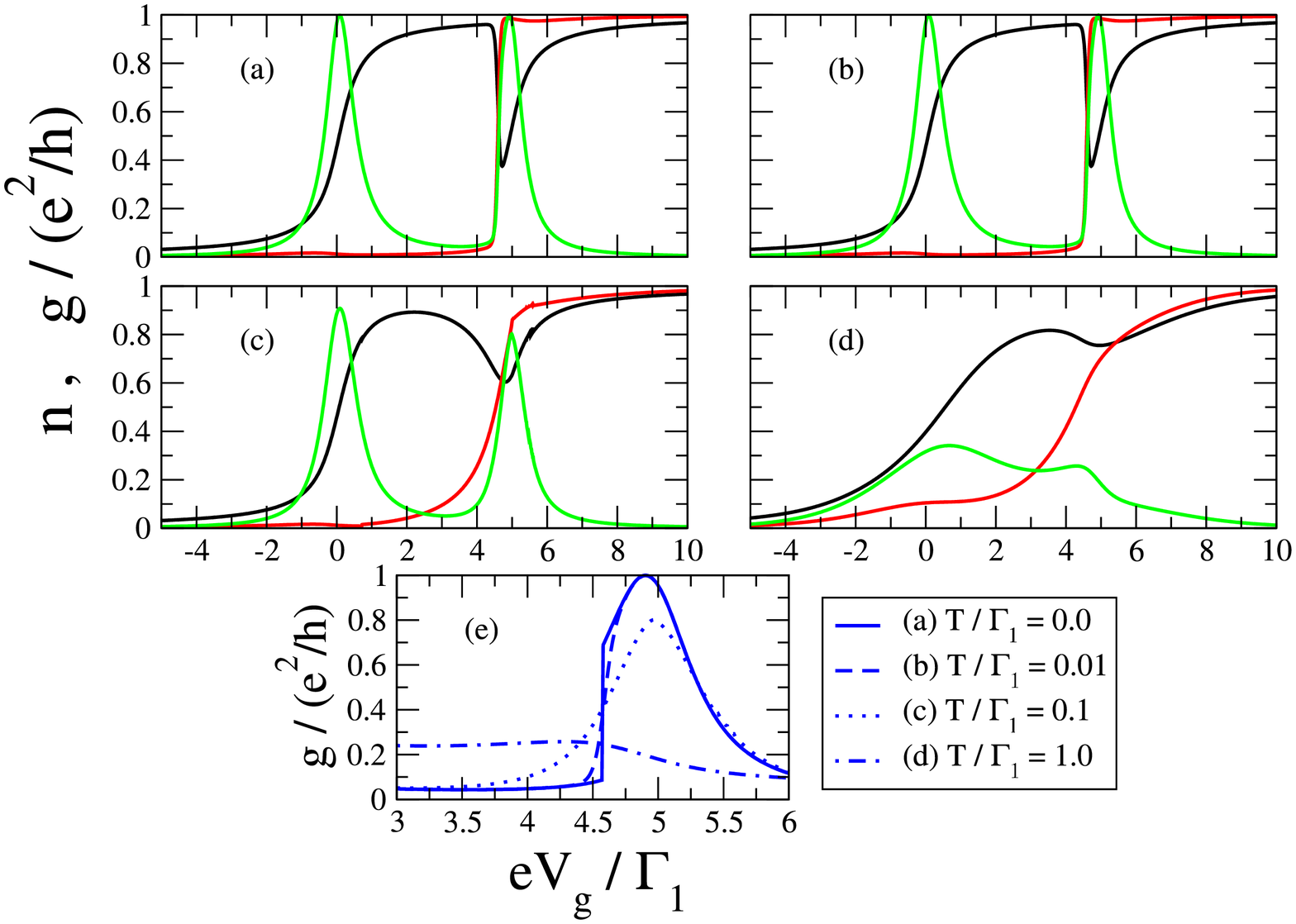}
\caption{\label{fig:temp} Level occupations (level 1 - black line; level
2 - red line) and conductance (green line) of a two-level dot
at various temperatures. The two levels are connected in the minus
configuration.
In all the graphs, $\epsilon_1/\Gamma_1=0.0$, $\epsilon_2/\Gamma_1=0.3$,
$\Gamma_2/\Gamma_1=0.2$, and $U/\Gamma_1=5.0$
(a) $T/\Gamma_1=0.0$;
(b) $T/\Gamma_1=0.01$;
(c) $T/\Gamma_1=0.1$;
(d) $T/\Gamma_1=1.0$;
(e) shows the conductance curves in the different temperatures together.}
\end{figure*}

Finally we remark on the effects of finite temperatures. In
Fig.~\ref{fig:temp} we examine one particular case (having parameter
values intermediate between those of Fig.~\ref{fig:mww}(c) and
Fig.~\ref{fig:mww}(d)). For temperatures lower than the narrow level
width, the only effect of the temperature is to make the discontinuity
smooth. Higher temperatures cause the smearing of the entire curves,
and the lowering of the conductance peaks heights below the maximal value
of $e^2/h$.

Thus, discontinuities may appear experimentally as fast gate voltage
dependence of the physical parameters, which thus show temperature
dependence for temperatures much lower than the, e.g., conductance peaks
widths. Of course, one cannot differentiate experimentally between
discontinuous features and continuous features having width smaller
than the lowest accessible temperature.

\section{\label{sec:conclude} Conclusions}

In this paper we examined some of the various phenomena which may
occur in a quantum dot where both interaction effects and inter-level
interference effects are important. We now turn to summarize our
results.

We have first shown that the occurrence of more conductance peaks
than energy levels in the dot can happen even in the non-interacting
case, and may help explain recent experimental observations.

In the interacting case we have found that in the minus configuration,
where the interference between the two levels does not affect their
populations, discontinuities occur if the levels are close enough,
the interaction is strong enough, and the ratio of the level widths
is much different from unity.
In contrast, in the plus configuration interference effects are
important. In the strong interaction regime discontinuities
can only occur when the levels are not too far away but not too
close; again, the widths ratio needs to be significantly different
from one.
In the weak interactions regime, the fast variation of the
population in the non-interacting case for almost degenerate levels
can lead to discontinuities in the interacting case even for
comparable widths. These results agree with the ``phase diagram''
picture of Golosov and Gefen \cite{golosov06} in the overlapping
part of the parameter space.

In both configurations, even when no discontinuity occurs, there is
usually a continuous version of mechanism~\ref{itm:mech_a} of
non-monotonous filling, accompanied by a broadening of the narrow
conductance peak. In addition, if the levels are not too far
apart, the continuous or discontinuous type~\ref{itm:mech_a}
non-monotonicity is accompanied by type~\ref{itm:mech_b}
non-monotonous behavior, causing asymmetry of the conductance peaks.

In either the plus or minus case, when no discontinuity occurs, or
when the discontinuity is weak enough, there is a conductance zero
in a location similar to the
non-interacting case. Strong enough discontinuities can cause the
conductance zero to be skipped.

Finite temperatures smear the discontinuities. The latter leave their
mark as sharp features in the gate voltage dependence of the different
physical properties, which show very strong temperature dependence
relative to other parts of the gate voltage dependence curves.

We conclude with a final remark. Our result in the interacting case
were obtained using the self-consistent Hartree-Fock approximation,
which neglects correlation effects. These may reduce the parameter
space regime in which discontinuities occur, or even eliminate it
completely. It can be expected, however, that the various continuous
behaviors found will remain even in a more complete theory.

\begin{acknowledgments}

We would like to thank Y. Gefen and D. Golosov for useful discussions.
Financial support from the Israel Science Foundation (Grant 877/04) is
gratefully acknowledged.

\end{acknowledgments}

\end{document}